\documentclass[12pt]{article}

 \newread\testifexists
 \def\GetIfExists #1 {\immediate\openin\testifexists=#1
     \ifeof\testifexists\immediate\closein\testifexists\else
     \immediate\closein\testifexists\input #1\fi}

 \usepackage{gthstyle}\usepackage{amsfonts}
 \usepackage{amssymb}
 \immediate\openin\testifexists=e
     \ifeof\testifexists\immediate\closein\testifexists\else
     \immediate\closein\testifexists\input e\fipsf

 \def\Bbb#1{\setbox0=\hbox{$\tt #1$} \copy0\kern-\wd0\kern .1em\copy0}

 \def\bbf#1{\setbox0=\hbox{$#1$} \kern-.025em\copy0\kern-\wd0
         \kern.05em\copy0\kern-\wd0 \kern-.025em\raise.0433em\box0}

 \immediate\openin\testifexists=a
     \ifeof\testifexists\immediate\closein\testifexists\else
     \immediate\closein\testifexists\input a\fimssym.def 

 \def\a{\alpha} \def\b{\beta}  
  \def\D{\Delta} \def\e{\varepsilon}

   \def\r{\varrho} \def\s{\sigma}

 \def\HH{\mathcal{H}}  \def\OO{{\cal O}} 
  \def\ra{\rightarrow}
  
   \def\ket{\rangle}

 \def\deff{\ {\buildrel{\rm def}\over{=}}\ }
 \def\iss{\ =\ }

 \def\fract#1#2{{\textstyle\frac{#1}{#2}}}
 \def\ffract#1#2{\raise .2 em\hbox{$\scriptstyle#1$}\kern-.3em/
                 \kern-.2em\lower .15 em \hbox{$\scriptstyle#2$}}
 
 \def\half{\fract12} \def\quart{\fract14} 
 
 \def\part#1#2{\frac{\partial#1}{\partial#2}}

   \newcommand{\fn}{\footnote}
 \newcommand{\nn}{\nonumber\\[2pt]} \newcommand{\nm}{\nonumber}
 \newcommand{\be}{\begin{eqnarray}} \newcommand{\ee}{\end{eqnarray}}
 \newcommand{\bi}[1]{\begin{itemize}\item[#1]} 
       \newcommand{\ei}{\end{itemize}}
 \newcommand{\eqn}[1]{(\ref{#1})}
 \def\bpm {\begin{pmatrix}}\def\epm{\end{pmatrix}}

 \newcommand{\crlb}[1]{\label{#1}\\[2pt]}
 \newcommand{\eela}[1]{\quad\hbox{\scriptsize{#1}}\label{#1}\end{eqnarray}}
 \newcommand{\eelb}[1]{\label{#1}\end{eqnarray}}
 
 \newcommand{\newsecb}[2]{\section{#1}\label{#2}\setcounter{equation}{0}}
 
 \newcommand{\nolabels} {\def\eel{\eelb} \def\crl{\crlb} \def\newsecl{\newsecb}}

\newcommand\publishversion{\nolabels\setlength{\textheight}{9in}\setlength{\oddsidemargin}{0in}
    \setlength{\textwidth}{6.3in}\setlength{\topmargin}{-0.1in}}

\publishversion 
\begin{document} \begin{titlepage}

\title{\normalsize \hfill ITP-UU-09/77 \\ \hfill SPIN-09/30
\\ \bigskip \bf{\Large
ENTANGLED QUANTUM STATES IN \\[5pt]A LOCAL DETERMINISTIC THEORY}}

\author{Gerard 't~Hooft}
\date{\normalsize Institute for Theoretical Physics \\
Utrecht University \\ and
\medskip \\ Spinoza Institute \\ Postbox 80.195 \\ 3508 TD
Utrecht, the Netherlands \smallskip \\ e-mail: \tt g.thooft@uu.nl \\ internet: \tt
http://www.phys.uu.nl/\~{}thooft/}

\maketitle

\begin{quotation} \noindent {\large\bf Abstract } \medskip \\ {\small{
Investigating a class of models that is familiar in studies of cellular automata, we
find that quantum operators can be employed to describe their long distance behavior. These
operators span a Hilbert space that appears to turn such a model into a genuine quantum field
theory, obeying the usual conditions of locality in terms of its quantum commutators. Entangled
states can be constructed exactly as in quantum theories.

This raises the question whether such models allow Bell's inequalities to be violated. Being a
local, deterministic theory, one would argue that this is impossible, but since at large distance
scales the model does not seem to differ from real quantum field theories, there is reason to
wonder why it should not allow entangled states. The standard arguments concerning Bell's
inequalities are re-examined in this light. }}
\end{quotation}

\flushleft{\footnotesize{Version August 24, 2009}}

\end{titlepage}

\eject

\newsecl{Introduction}{intro}

Local hidden variable theories are usually formulated in such a way that Bell's inequalities for
experiments with entangled particles can be derived, as soon as these particles are spacelike
separated. Since quantum-entangled particles violate Bell's inequalities, while they can easily be
spacelike separated, it is generally concluded that no local hidden variable theory can exist that
reproduces typical quantum phenomena. In this paper, however, we show that deterministic models can be described in a way not usually
considered. 

Quantum mechanics would have been an impeccable theory for nature's dynamical degrees of freedom at the tiniest distance scale, if the gravitational
force would not have required invariance under general coordinate transformations. Today's best candidate for a generally covariant quantum theory is
superstring theory, but this theory still appears to exhibit important gaps and it is difficult to interpret at a more fundamental level. Because of
a gigantic landscape of self-consistent solutions, that theory also does not produce significant predictions.

The difficulty with quantum gravity appears to be a consistent description of Hilbert space. This difficulty becomes manifest in
one of the simplest generally covariant cosmological models: gravitating particles in 2 spacelike and 1 timelike dimensions.
Without quantum mechanics, a closed universe with genus \(g\) and \(N\) gravitating particles has \(4N+12(g-1)\) freely
adjustable parameters: \(2N\) for the positions, \(2N\) for the momenta, while the gravitational degrees of freedom, \(12(g-1)\)
of them, are only topological (when \(g=0\), they produce 12 \emph{constraints} on the other parameters). The theory has Poisson
brackets\cite{gth2d}, and one would be tempted to replace these by commutators, expecting that this would generate a quantum
mechanical Hilbert space\cite{carlipea}. However, not all is well. Attempts to characterize this Hilbert space, find its spectrum
of states, and to interpret the Hamiltonian as a generator of a unitary evolution law seem to fail, precisely because the
coordinates and momenta can only be defined consistently in a classical topological background, which is what we do not have in
quantum mechanics. In this theory, it is as if we cannot discard the classical topological definitions; the holonomies refer to
the conventional local Poincar\'e group, and the commutation rules for its generators remain as in the classical theory. Since
positions and momenta of particles must be associated to holonomies, it seems to be impossible to demand that these should obey
quantum commutation rules. This model is the most concise cosmological model we have, and it seems to refuse being
quantized.\fn{The successes reported in Ref.~\cite{carlipea} mainly refer to the topological degrees of freedom of the pure
gravity sector, which are non-local. It is the local, particle degrees of freedom that cause difficulties that have not yet been
adequately addressed, in particular the causality and unitarity issues in the case of a closed universe.}

After careful examination of this difficulty, we could rephrase it as being an obstacle against any attempt to formulate the quantum version of a
compact cosmological system. What would the role of an observer be, and how do we define statistics here? In this context, it seems that Einstein's
original objections against quantum mechanics are justified. In a finite universe, something happens or it does not; one does not have the
opportunity to do repeated experiments so as to make statistical statements. In a finite universe, quantum mechanics may only refer to
approximations, not to an exact theory.

Even in a finite universe, one may assume the approximation that a system is surrounded by an asymptotically flat and featureless vacuum. One might
suspect that, what we call quantum mechanics today, is the statistical description of any part of the universe for which this approximation is
meaningful. Thus one may assume that the vacuum itself is actually a complicated solution of the evolution equations, allowing a precisely formulated
statistical description that is indeed different from the statistical description of other states, such as the one-particle state, the 2 particle
state, etc. This, we shall illustrate in a class of models that we call cellular automata.\fn{The word ``cellular automata" is the plural of
``cellular automaton", which would be one particular choice among our models.}

Cellular automata are ideally suited for simulations on a computer, even if we assume our models to be defined on an infinite
size lattice. However, unlike a computer model, we can consider states defined as quantum superpositions of individual states as
generated by a computer. This leads to an observation that we wish to underline: the \emph{evolution law} may be chosen to be
entirely classical, but this does not forbid the use of \emph{states} that are quantum superpositions. In our way of approaching
these models mathematically, one quickly reaches the stage where the evolution of these quantum states is studied, operators are
defined to act on these states, while the direct connection to the `ontological' states of the original model might get lost.
This typically happens as one performs scale transformations to address large-distance features. Small-distance features are far
too detailed; we loose those details out of sight, and continue to work with a small subspace of the original Hilbert space,
typically the subspace spanned by the lowest energy eigenstates of the time evolution operator \(H\). All states that we keep are
highly entangled ones, and subsequently one of the things we also loose control over are the Bell inequalities.\cite{JB}

This argument would not be good enough to dismiss the Bell inequalities altogether. We can reconstruct them, simply by realizing that the underlying
model still is a deterministic cellular automaton, and this is why we shall try to analyze the Bell inequalities as precisely as we can: if Bell
inequalities still apply to our system, are they indeed the ones that are routinely being tested in many of today's carefully designed experiments,
or are they formal features that have nothing to do with presently observable quantities? This is the important question that we wish to address.

Our procedure requires two ingredients: advanced mathematics and delicate physical requirements. The mathematics we use will seem
to be flawed: we employ an expansion (the Baker-Campbell-Hausdorff expansion) that may well be fundamentally divergent in the
case of interest. We then use physical considerations to argue that, nevertheless, the amputated series should be meaningful.
Other physical observations are the numerous symmetries that we encounter in our world and that seem to be entirely unbroken:
rotation, translation and Lorentz invariance, for instance. These symmetries do not hold at all in our models, and we admit that
it seems to be prohibitively difficult to implement these symmetries in a satisfactory way. But then one could argue that
symmetry arguments should not enter into the discussion of the interpretation of quantum mechanics. Of course, the author has
attempted to theorize about the physical origin of these symmetries, and we think they can be explained, but most of this would
go beyond the scope of this paper.

As for our ``flawed" mathematics, we already indicated that there may be circumstances where the divergences cancel or can be cured; again, we bring
forward that this question may have little to do with the logical observation that quantum superpositions of states can be important ingredients of a
theory even if its fundamental evolution law can be considered to be a classically deterministic one, if the ingredients needed to phrase this
classical law contain elements that are far beyond what can be observed today. Admittedly, this sounds very much as a hidden variable theory, but
these hidden variables are far from the concocted ones that we see in the more well-known representatives of the hidden variable theories, and in our
models no need is seen to postulate them to be non-local.

A pseudo-non local feature can be built in our model: we suspect that it may be necessary to allow for ``information loss" to
occur at the level of the cellular automaton evolution laws themselves. Information loss here simply means that evolution takes
place in the forward-time direction; evolution ``backwards" to the past may well be impossible, just because many different
states at \(t=t_0-1\) might evolve into the same state at \(t=t_0\). This would force us to rephrase our model in terms of
\emph{equivalence classes} that should now take the place of elementary basis elements of Hilbert space. Two distinct ontological
states at \(t=t_1\) would be considered equivalent if, at a later time \(t=t_1+T\), they both have evolved into the same state.
These classes would induce a considerable amount of complexity at the elementary level of our models; only in the simplest of all
cases our equivalence classes may coincide with, for instance, equivalence under local gauge transformations. Still, there would
be no ``communication at a distance", exactly in accordance with the situation we have in all those successful quantum field
theories that describe the experimentally observed standard model interactions.


\newsecl{Standard Quantization and Primitive Quantization}{prequant}

The standard quantization procedure is a method to produce a quantum model starting from a classical mechanical model. One considers the Poisson
brackets of the classical model and replaces these by commutator expressions. The quantum model thus obtained is physically very different from the
original classical model, but the classical model does correspond to the limit \(\hbar\ra 0\) of the quantum model, and the procedure has the
important feature that, in most cases in practice, it is unique, so we have a one-to-one mapping of classical systems onto quantum systems.

Here, however, we emphasize that there exists a very different way to obtain a quantum theory starting from a classical one. In what we will refer to
as the \emph{primitive quantization procedure}\fn{Also named \emph{pre-quantization} in previous publications of this author\cite{GtHbeable}.}, we
actually make no changes at all in the physical features of the model. Physically, the classical model remains classical, but we do introduce
operators as if the model were a quantum theory.

Even in a theory based on classical logic, such as Newtonian mechanics, one can ask the following question: \emph{how would a tiny change in our
dynamical degrees of freedom propagate in time?} Such a tiny change would be indicated in terms of an operator that does not commute with the
operators that describe the dynamical variables themselves. All these operators could possess important symmetry relations with respect to one
another, and they could all be important for understanding the system.

For our future considerations, in any deterministic system, we now introduce three different kinds of observable operators: \emph{beables},
\emph{changeables}, and \emph{superimposables}.

A beable \(B\) is an operator that does not affect the state a system is in, but measures some of its ontological
properties\cite{GtHbeable}. All beables, at all times \(t\), will always commute with one another; a basis exists, called the
\emph{ontological basis}, in which all these operators are diagonal at all times. In the well-known Standard Model, without any
changes, a significantly interesting set of beables cannot be found, but it might be possible to find one at the Planck scale.
Also, large sets of beables can be found for non-interacting Bose fields\cite{bosons} and for massless fermions\cite{fermions}.

A changeable is any operator that maps an ontological state onto another ontological state. The time evolution operator, and many of the symmetry
operators are changeables. Thus, in general, a changeable maps a cellular automaton onto any other cellular automaton, which could be the original
one displaced in space and time, for instance. If \(B\) is a beable and \(C\) is a changeable, we have an other beable \(B'\) such that
    \be B'C=CB\ . \eel{changebe}

A superimposable \(S\) is an operator that maps an ontological state onto a superposition of different ontological states; the generic observable
operator is a superimposable. In an ontological basis, the three types of operators can be characterized as
 \be B={\scriptstyle\pmatrix{*\quad\quad\quad\quad\cr\quad*\quad\quad\quad\cr\quad\quad*\quad\quad\cr\quad\quad\quad*\quad\cr\quad\quad\quad\quad*}}\ ,\qquad
     C={\scriptstyle\pmatrix{\quad*\quad\quad\quad\cr\quad\quad\quad*\quad\cr*\quad\quad\quad\quad\cr\quad\quad\quad\quad*\cr\quad\quad*\quad\quad}}\ ,\qquad
     S={\scriptstyle\pmatrix{*\,*\,*\,*\,*\cr*\,*\,*\,*\,*\cr*\,*\,*\,*\,*\cr*\,*\,*\,*\,*\cr*\,*\,*\,*\,*}} ,\eel{BCSchar}
where \(*\) stands for any non-vanishing matrix element.

Changeables and superimposables are not diagonal in the ontological basis, and so they need not commute with most of the beables and other
observables.

Just as in any quantum theory, we can describe the time evolution of any deterministic system in terms of a hamiltonian. For infinitesimal time
shifts, we need a superimposable. Finite time evolutions are described by changeables. At first sight, one obtains sets of equations that appear to
be very similar to those describing a fundamental quantum theory. Often, however, one encounters important differences. In particular, the
hamiltonian usually does not seem to have a ground state, so that it cannot be used to do thermodynamics, but, as we will see, there are cases where
this difficulty might be cured.

Thus, theories obtained by primitive quantization form a subset of all quantum theories. They are characterized by the fact that \begin{quotation}
\noindent a non-trivial set of operators exists, called `beables', that describe the system at all times, while they all commute with one
another.\end{quotation} \noindent The basis in which all beables are diagonal is called the `ontological basis'.

\newsecl{The Cellular Automaton Model}{CA}

The model described in this Section will at first sight seem to do exactly what we want: there are only beables evolving into beables at a
microscopic level, whereas something resembling a full-fledged quantum field theory emerges at large scales. Indeed, we have a hamiltonian with a
natural lower bound, so that a physical vacuum state exists. The caveats however will come at the end.

Our cellular automaton is a totally classical model, and it is completely deterministic and local in the classical sense, but its large scale
features are so complex that they require quantum mechanical techniques for addressing their statistics. Thus, we introduce beables and changeables
\emph{as if} we were dealing with a quantum theory. The model considered in this paper will be time-reversible\cite{Fredkin}\fn{As stated in the
introduction, Section \ref{intro}, more realistic models representing the real world are likely to be time irreversible\cite{disdet}.}.

Space and time\cite{balachandran} are both taken to be discrete: we have a \(D\) dimensional space, where positions are indicated
by integers: \(\vec x=(x^1,\,x^2,\,\cdots,\,x^D)\), where \(x^i\in\Bbb Z\). Also time \(t\) will be indicated by integers, and
time evolution takes place stepwise. The physical variables \(F(\vec x,\,t)\) in the model could be assumed to take a variety of
forms, but the most convenient choice is to take these to be integers modulo some number \(\Bbb N\). We now write down an
explicit model, where these physical degrees of freedom are defined to be attached only to the even lattice sites:
 \be \sum_{i=1}^D x^i+t\iss\ \hbox{even}. \eel{evensites}
Furthermore, the data can be chosen freely at two consecutive times, so for instance at \(t=0\), we can choose the initial data to be \(\{F(\vec x,\
t=0),\ F(\vec x,\ t=1)\}\).

The dynamical equations of the model can be chosen in several ways, provided that they are time reversible. To be explicit, we choose them to be as
follows:
 \be && F(\vec x,\,t+1)\iss F(\vec x,\,t-1)+\nn
 && Q\left(F(x^1\pm 1,x^2,\cdots,x^D,\,t),\,\cdots,\,F(x^1,\cdots,x^D\pm
 1,\,t)\right)\ \hbox{Mod }\Bbb{N}\ , \label{caeq}\\
 &&\hbox{when}\quad\textstyle{\sum_i x^i+t}\quad\hbox{is odd}\,,\nm\ee
where the integer \(Q\) is some arbitrary given function of all variables indicated: all nearest neighbors of the site \(\vec x\) at time \(t\). This
is time reversible because we can find \(F(\vec x,\,t-1)\) back from \(F(\vec x,\,t+1)\) and the neighbors at time \(t\). Assuming \(Q\) to be a
sufficiently irregular function, one generally obtains quite non-trivial cellular automata this way. Indeed, this category of models have been shown
to contain examples that are computationally universal \cite{Fredkin}. Models of this sort are often considered in computer animations.

We now discuss the mathematics of this model using Hilbert space notation. We switch from the Heisenberg picture, where states are fixed, but
operators such as the beables \(F(\vec x,\,t)\) are time dependent, to the Schr\"odinger picture. Here, we call the operators \(F\) on the even sites
\(X(\vec x)\), and the ones on the odd sites \(Y(\vec x)\). As a function of time \(t\), we alternatingly update \(X(\vec x)\) and \(Y(\vec x)\), so
that we construct the evolution operator over two time steps. Keeping the time parameter \(t\) even:
 \be U(t,\,t-2)=A\cdot B\ , \eel{evolv}
where \(A\) updates the data \(X(\vec x)\) and \(B\) updates the data \(Y(\vec x)\).
\def\PP{\mathcal{P}}

\def\AA{a} \def\BB{b}
Updating the even sites only, is an operation that consists of many parts, each defined on an even coordinate \(\vec x\), and all commuting with one
another:
 \be A=\prod_{\vec x\ \mathrm{even}} A(\vec x)\ ,\quad[A(\vec x),\,A(\vec x\,')]=0\ , \eel{Aprod}
whereas the \(B\) operator refers only to the odd sites,
 \be B=\prod_{\vec x\ \mathrm{odd}} B(\vec x)\ ,\quad[B(\vec x),\,B(\vec x\,')]=0\ . \eel{Bprod}
Note now, that the operators \(A(\vec x)\) and \(B(\vec x\,')\) do not all commute. If \(\vec x\) and
\(\vec x\,'\) are neighbors, then
 \be \vec x-\vec x\,'=\vec e\ ,\quad |\vec e\,|=1\quad \ra\quad [A(\vec x),\,B(\vec x\,')]\ne 0\ .
 \eel{ABnoncomm}
It is important to observe here that both the operators \(A(\vec x)\) and \(B(\vec x)\) only act in finite subspaces of Hilbert space, so we can
easily write them as follows:
 \be A(\vec x)=e^{-i\AA(\vec x)}\ , \qquad B(\vec x)=e^{-i\BB(\vec x)}\ . \eel{ABexp}
Note that \(A(\vec x)\) and \(B(\vec x)\) are changeables, while \(\AA(\vec x)\) and \(\BB(\vec x)\) will be superimposables, in general. We can
write
 \be\AA(\vec x)= \PP_x(\vec x)\ Q(\{Y\})\ , \quad\BB(\vec x)= \PP_y(\vec x)\ Q(\{X\})\ , \eel{steps}
where \(\PP_x(\vec x)\) is the generator for a one-step displacement of \(X(\vec x)\):
 \be e^{i\PP_x(\vec x)}|X(\vec x)\ket\deff |X(\vec x)-1\ \hbox{Mod }\Bbb{N}\ket\ , \eel{stepdef}
and, similarly, \(\PP_y(\vec x)\) generates one step displacement of the function \(Y(\vec x)\). We see that
 \be &&[\AA(\vec x),\,\AA(\vec x\,')]=0\ ,\quad[\BB(\vec x),\,\BB(\vec x\,')]=0\ ,\quad\forall\,(\vec
x,\,\vec x\,')\ ; \crl{AABBcomm} &&[\AA(\vec x),\,\BB(\vec x\,')]=0\quad \hbox{only if}\quad|\vec
x-\vec x\,'|>1\ . \eel{ABnonc}

A consequence of Eqs.~\eqn{AABBcomm} is that also the products \(A\) in Eq.~\eqn{Aprod} and \(B\) in Eq.~\eqn{Bprod} can be written as
 \be A=e^{-i\sum_{\vec x\ \mathrm{even}}\AA(\vec x)}\ , \qquad B=e^{-i\sum_{\vec x\ \mathrm{odd}}\BB(\vec
x)}\ . \eel{sumAB}
 However, now \(A\) and \(B\) do not commute. Nevertheless, we wish to compute the total evolution
operator \(U\) for two consecutive time steps, writing it as
  \be U=A\cdot B=e^{-i\AA}\,e^{-i\BB}=e^{-2iH}\ . \eel{UexpH}
For this calculation, we could use the power expansion given by the Baker-Campbell-Hausdorff formula,
 \be && e^P\,e^Q=e^R\ ,\nn
 && R=P+Q+\half[P,Q]+\fract{1}{12}[P,[P,Q]]+\fract 1{12}[[P,Q],Q]+\fract{1}{24}[[P,[P,Q]],Q]+\cdots, \nn \eel{BCH}
a series that continues exclusively with commutators\cite{SW}. Replacing \(P\) by \(-i\AA\), \(Q\) by \(-i\BB\) and \(R\) by \(-2iH\), we find a
series for the `hamiltonian' \(H\) in the form of an infinite sequence of commutators. Now note that the commutators between the local operators
\(\AA(\vec x)\) and \(\BB(\vec x\,')\) are non-vanishing only if \(\vec x\) and \(\vec x\,'\) are neighbors, \(|\vec x-\vec x\,'|=1\). Consequently,
if we insert the sums \eqn{sumAB} into Eq.~\eqn{BCH}, we obtain again a sum:
 \be && H=\sum_{\vec x}\HH(\vec x)\ ,\nn
 &&\HH(\vec x)=\half\AA(\vec x)+\half\BB(\vec x)+\HH_2(\vec x)+\HH_3(\vec x)+\cdots\ ,
 \eel{Hsum} where
\be \HH_2(\vec x)&=&- \fract 14 i\sum_{\vec y}[\AA(\vec x),\,\BB(\vec y)]\ , \nn
 \HH_3(\vec x)&=&-\fract 1{24} \sum_{\vec y_1,\,\vec y_2} [\AA(\vec x)-\BB(\vec x)\,,\ [\AA(\vec
 y_1),\BB(\vec y_2)]] \ , \quad\hbox{etc.} \eel{Hdensitysum}
All these commutators are only non-vanishing if the coordinates \(\vec y\), \(\vec y_1\), \(\vec y_2\), etc., are all neighbors of the
coordinate \(\vec x\). It is true that, in the higher order terms, next-to-nearest neighbors may enter, but still, one may observe that
these operators are all local functions of the `fields' \(F(\vec x,\,t)\), and thus we arrive at a hamiltonian \(H\) that can be regarded
as the sum over \(D\)-dimensional space of a Hamilton density \(\HH(\vec x)\), which has the property that
 \be[\HH(\vec x),\,\HH(\vec x\,')]=0\ ,\quad\hbox{if}\quad|\vec x-\vec x\,'|\gg 1\ .
\eel{Hdensitycomm} At every finite order of the series, the Hamilton density \(\HH(\vec x)\) is a finite-dimensional Hermitean
matrix,\fn{Actually, we are only interested in the conjugacy classes of \(H\). Writing in Eq.~\eqn{BCH}, \(P=S+D,\quad Q=S-D\), the
conjugacy classes for \(R\) are characterized by terms that are odd in \(S\) and even in \(D\). Indeed, one can find an expansion
\(F=-\half D+\fract1{24}[S[S,D]]+\cdots\) such that, in Eq.~\eqn{BCH}, we have
\(e^F\,R\,e^{-F}=2S-\fract1{12}[D[S,D]]+\fract1{960}(\,8\,[D[S[S[S,D]]]]-[D[D[D[S,D]]]]\,)+\OO(S,D)^7\), which is even in \(D\) and odd in
\(S\).} and therefore, it will have a lowest eigenvalue \(h\). In a large but finite volume \(V\), the total hamiltonian \(H\) will
therefore also have a lowest eigenvalue, obeying
 \be E_0>h\,V\ . \eel{eigenH}
The associated eigenstate \(|0\ket\) might be identified with the `vacuum'. This vacuum is stationary, even if the automaton itself may
have no stationary solution. The next-to-lowest eigenstate may be a one-particle state. In a Heisenberg picture, the fields \(F(\vec x,t)\)
may create a one-particle state out of the vacuum. Thus, we arrive at something that resembles a genuine quantum field theory. The states
are quantum states in complete accordance with a Copenhagen interpretation. The fields \(\AA(\vec x,t)\) and \(\BB(\vec x,t)\) should obey
the Wightman axioms.

There are three ways, however, in which this theory differs from conventional quantum field theories. One is, of course, that space and
time are discrete. Well, maybe there is an interesting `continuum limit', in which the particle mass(es) is(are) considerably smaller than
the inverse of the time quantum.

Secondly, no attempt has been made to arrive at Lorentz invariance, or even Gallilei invariance. Thus, the dispersion relations for these
particles, if they obey any at all, may be nothing resembling conventional physical particles. Do note, however, that no physical
information can travel faster than velocity one in lattice units. This is an important constraint that the model still has in common with
special relativity.

But the third difference is more profound. It was tacitly assumed that the Baker-Campbell-Hausdorff
formula converges. This is often not the case. In Section~\ref{convergence}, we argue that the
series will converge well only if sandwiched between two eigenstates \(|E_1\ket\) and \(|E_2\ket\)
of \(H\), where \(E_1\) and \(E_2\) are the eigenvalues, that obey
 \be 2|E_1-E_2|< 2\pi\hbar/\D t\ , \eel{eigenconstraint}
where \(\D t\) is the time unit of our clock, and the first factor 2 is the one in Eq.~\eqn{UexpH}. (``Planck's constant", \(\hbar\), has been
inserted merely to give time and energy the usual physical dimensions.)

This may seem to be a severe restriction, but, first, one can argue that \(2\pi\hbar/\D t\) here is the Planck energy, and in practice, when we do
quantum mechanics, we only look at energies, or rather energy differences, that indeed are much smaller than the Planck energy. Does this mean that
transitions with larger energy differences do not occur? We must realize that energy is perhaps not exactly conserved in this model. Since time is
discrete, energy at first sight seems to be only conserved modulo \(\pi\), and this could indicate that our `vacuum state' is not stable after all.
The energy might jump towards other states by integer multiples of \(\pi\). In Section \ref{convergence} however, we argue that such violations of
energy conservation will \emph{not} occur, and the existence of an hamiltonian density is a more profound property of all cellular automata that
allow time reversal (so that the evolution is obviously unitary).

The conclusion we are able to draw now, is that procedures borrowed from genuine quantum mechanics can be considered, and they may lead to a
rearrangement of the states in such a way that beables, changeables and superimposables naturally mix, leaving an effective description of a system
at large time and distance scales for which only quantum mechanical language applies. This, we think, is all we really need to understand why it is
quantum mechanics that seems to dominate the world of atoms and other tiny particles, which, though small compared to humans, are still very large
compared to the Planck scale.

\newsecl{The Bell inequalities.} {Bell}

\begin{figure}[h] \begin{quotation}
 \epsfxsize=125 mm\epsfbox{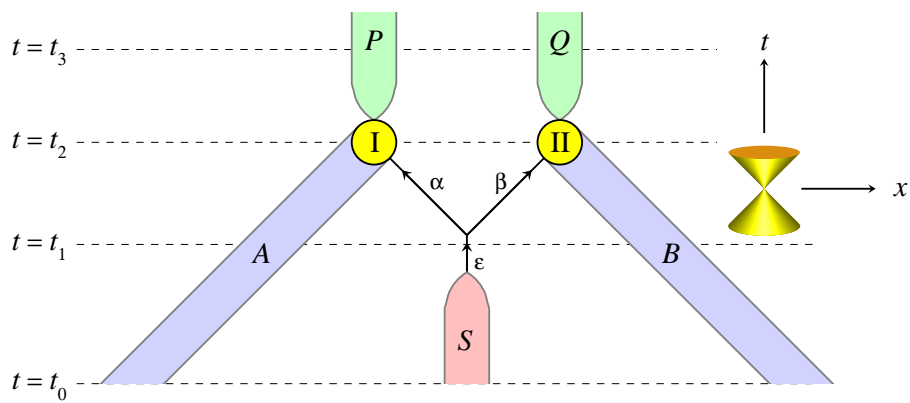}
  \caption{Space-time diagram of an experiment with entangled particles \(\a\) and \(\b\).
Coordinates $x$ and $t$ are chosen such that the speed of light is \(45^\circ\), giving a light
cone as displayed. Classical data are displayed here in wide, colored trajectories, while quantum
objects that may be entangled are indicated with single lines. Explanation of symbols: see text.}
  \label{quantumexperiment.fig}\end{quotation}
\end{figure}

We now sketch the prototype of the \emph{Gedanken} experiments that are commonly used\cite{Seevinck}\cite{CK} to prove that no
deterministic theory can reproduce the situations that can occur in quantum theories as sketched in the previous section, using
Figure~\ref{quantumexperiment.fig}. 
A source \(S\) produces a quantum state \(\e\), for instance an unstable particle, that leads to two entangled objects \(\a\) and \(\b\). The objects
are separated, at velocities close to that of light, and then both measured, by Alice (\(I\)) and Bob (\(II\)), who are spacelike separated. Alice
and Bob choose between two or more non-commuting observables to measure.

In a completely deterministic model, Alice and Bob cannot have the ``free will" to modify the settings of their measurement apparatus without some
explicit change in their past --- that's what determinism is about. However, this observation by itself does not remove the paradox: the data that
affect Alice and Bob's decisions may still be spacelike separated. To be explicit, it is instructive to assume their decisions to depend on signals
that reach them from opposite directions. These signals could be \emph{classical} signals from, say, two far away quasars.\fn{Quasars are known to be
fluctuating light sources; these fluctuations could be used to make decisions. We might use either classical or quantum fluctuations.}

In Figure~\ref{quantumexperiment.fig}, the trajectories of typical quantum states are indicated by single lines, while macroscopic data, which
generally occupy more space, are denoted by wider colored bands. The quasar chosen by Alice leads to a setting \(A\), and the outcome of her
measurement is \(P\), while Bob finds a setting \(B\), leading to an outcome \(Q\). The fact that \(\a\) and \(\b\) have a quantum entanglement
implies that the probability \(W(P,Q|A,B)\) for these outcomes cannot be written as a product \(W_1(P|A)\,W_2(Q|B)\) or a convolution of such products, \(\sum_i\r_iW_1^i(P|A)\,W_2^i(Q|B)\).
Yet \(P\) and \(B\) are entirely spacelike separated, and so are \(A\) and \(Q\). If \(\a\) and \(\b\) were \emph{classical} objects,
accompanied by no matter how many `hidden variables', some `spooky signal' had to be sent from \(A\) to \(Q\) and from \(B\) to \(P\). This would
violate special relativity, or it would require some form of non-locality. This kind of non-locality is usually dismissed by most authors.

However, one may still ask whether this really proves the nonexistence of an ontological evolution law. The entangled particles
\(\a\) and \(\b\) are evidently described by non commuting operators, but that concerns the description of the \emph{states} they
are in, and not the law according to which they \emph{evolve}. Very often, considerations of the consequences of Bell's
inequalities hardly mention the evolution law. How did the entangled particles form? Can entangled particles be produced by
objects that start out as `classical' states? Does the evolution have to involve operators that do not commute after some finite
time interval?

Consider now a small disturbance of quasar \(A\) at time \(t=t_0\). Such a small, localized disturbance may well affect Alice, persuading her to
rotate her measuring device, but it cannot affect Bob, nor the particles \(\e\),\(\a\) or \(\b\). If Alice could be persuaded to switch from one
operator to an other operator that does not commute with the first, we inevitably arrive at a contradiction. However, if the perturbation induced in
the quasar was one that modified only the ontological state the quasar is in, then this modification is described by a changeable, not by a
superimposable. This may well be essential. The quasar stays in an eigenmode of the beables. It can never make a transition into a superposition of
different ontological states.

If, on the other hand, Alice wishes to rearrange her measuring device from measuring one operator \(\s_1\) to measuring another
operator \(\s_2\) that does not commute with \(\s_1\), she does something that cannot be expressed in terms of beables alone. The
rotation of her device is described not by a changeable, but by a superimposable operator. Somewhere along the line,
superimposable operators must have come into play.

Therefore, Alice cannot make such a transition if the quasar had only been affected by a changeable operator. This particular
disturbance of the quasar is not of the right kind to cause Alice (or rather her measuring device) to go into a superimposed
state. From the state she is in, she cannot measure the new operator \(\s_2\).

Clearly, the only way to handle quantum entangled states, is by assuming that these states were quantum-entangled right from day
one in the universe. This, we think, should not be a problem at all.

A correct description of all statistical features of the universe should be such that, right from the beginning, the probability
distributions were best described by using the operators that we find useful today for describing the Standard Model. Those were
eigenstates that always have been entangled. Indeed, quasar \(A\) and quasar \(B\) have always been in a quantum entangled state.
Any perturbation that we would like to consider, would be easiest described by a superimposable operator, not just by a beable or
even a changeable.

The above now raises the following question. Consider a quantum field theory such as the Standard Model. Is there a basis containing time
dependent observables describing dynamical variables at the Planck scale, that are all in a diagonal form and are sufficiently dense in the
entire space of observables that we usually employ to do physics? In short: can we identify the beables? For instance, can we find such
beables that also represent classical configurations at large physical scales in an approximation that becomes exact in the macroscopic
limit? This is an interesting challenge for quantum field theories\cite{bosons}, or perhaps superstring theory\fn{quantized strings consist
mainly of exactly harmonic oscillators, for which interesting descriptions in terms of beables van be found\cite{bosons}.}, but we will
leave it unanswered here.

It is our conjecture that we can consider our cellular automaton model in two mathematically equivalent ways; on the one hand we
have the totally classical evolution equations, such as those of our cellular automaton, Eqs.~\eqn{caeq}, and on the other hand
we have their description in terms of the Schr\"odinger equation of the primitively quantized theory (Section~\ref{prequant}),
associated with a hamiltonian \(H\), see Eq.~\eqn{Hsum}. Is this contradicted by the Bell inequalities\cite{JB}? There may be
different ways to circumvent such a conclusion.

One is, as stated above, that a pure changeable opersator would replace the beables for quasar \(A\) into other beables, and thus
not allow Alice to turn her detector into the non-diagonal eigenstate needed to measure the new operator. We must then assume the
quasars to be in an entangled state from the start.

Alternatively, let us again first look at the events from a classical point of view, in terms of the beables. At \(t=t_0\) only a
small, classical, modification is introduced in quasar \(A\), described by a changeable, and suppose that it \emph{could} have
persuaded Alice to change her mind about the setting of her experiment at \(t=t_2\). Quasar \(B\) is not affected, so that Bob
does the same measurement as before, and also the atom \(\e\) at \(t=t_1\) is unaffected. The measurements are registered and the
classical registrations are compared at \(t=t_3\). It is here that we would encounter the usual contradiction: the particles
\(\a\) and \(\b\) cannot be in the quantum entangled state that we expect from the experiment at \(t=t_1\). If this is the case,
one must conclude that a classical perturbation in quasar \(A\) either forced the entangled particles \(\a\) and \(\b\) into an
unallowed state (a state contaminated with spin 2 states in the case of two photons), or it may have had a superluminal effect on
the surrounding particles, at least all the way to the system \(S\).

Why would a state such as the spin 2 state at \(t=t_1\) not be allowed? This state was merely not expected because of our physical
experiences in this universe; the experiment was expected to yield an entangled state, \emph{but our ``wrong" state is in no way a
forbidden state!} We can extrapolate this ``forbidden" state backwards in time, finding that particle \(\e\) is now replaced by a highly
improbable object with higher spin. Solving Schr\"odinger's equation further backwards in time, we see that the entropy rapidly increases,
yielding a state at \(t=t_0\) whose entropy is far greater than we usually assume for the far past. But, since we can also solve the
\emph{classical} equations backwards in time, we conclude that this highly improbable state must coincide with our classically perturbed
cellular automaton. We must conclude that the very high entropy state at \(t=t_0\) mathematically coincides with the state obtained when
the very low entropy configuration of the two quasars was slightly perturbed classically. We claim that such equivalences may occur if we
describe an ontological state using our hamiltonian \(H\) while inserting the wrong phase factors. The quantum states we use in our
effective quantum description strongly depend on these phases while on the other hand the phase factors in the wave functions are entirely
fictional. In short: there is no contradiction if we accept the possibility that a classical cellular automaton in a given classical state
can be described not only as a quantum field in a low entropy state, but it can equivalently be described by a quantum state with very high
entropy. The point is that, in our classical states, the phase factors are ill-defined. If we choose them ``incorrectly" it may easily
happen that the quantum state we use to describe it is totally different from the most preferred description.

In physics in general, including the physics of quantized fields, we have the second law of thermodynamics telling us that entropy
increases with time. If we extrapolate an arbitrary state in Hilbert space backwards in time, chances are that entropy will not decrease
back to its early value, and we arrive at a vast majority of states in Hilbert space that are physically unacceptable. Thus, we note that
in our preferred description of nature we actually choose our Hilbert space to be much smaller than the one spanned by all possible
ontological states. The phase factors are very special. Thus, the entangled states are needed to describe our universe from day one.
\emph{In any meaningful description of the statistical nature of our universe, one must assume it to have been in quantum entangled states
from the very beginning.} In contrast, the evolution law may always have been a deterministic one.

The reason why the entropy in the early universe was as low as we usually assume, is that the universe has expanded. Even though total
entropy continuously increases, the entropy density often decreases, since the cosmic background radiation temperature decreases in time.

The above considerations lead us to realize that the set of states we use to describe physical
events are characterized by two important extensive quantities: the \emph{total energy} and the
\emph{total entropy}. In all states that we consider, both these quantities are very small in
comparison with the generically possible values. Any ontological state of our automaton,
characterized by beables, is a superposition of all possible energies and all possible entropies.
The state we use to describe our statistical knowledge of the universe has very low energy and very
low entropy. It appears that, if in any ontological state we make one local perturbation, the
energy will not change much, but the entropy increases tremendously, thus allowing particles \(\a\)
and \(\b\) at \(t=t_1\) to enter into a modified, (dis)entangled state. If we perturb the quantum
state, both energy and entropy change very little, \(\a\) and \(\b\) stay in the same state, but
then Bell's inequality needs not be obeyed.

We conclude that Bell's inequalities might not invalidate our model, but imply that a local
perturbation of the beables (near quasar \(A\) in our example) cannot be associated with a local
perturbation of the quantum state that we use to describe the evolution of our universe. A local
classical perturbation is possible, but would have to be accompanied by huge modifications of the
total entropy, or alternatively changes in the phase factors all over space, so that the only
quantum state with the same low total entropy would contain non-local modifications

It seems that the clue to the resolution of the Bell inequality paradox ensuing from our model is
that the quantum phase factors that relate our quantum states to the ontological states of the
theory, are fundamentally unobservable, since, indeed, they are ill-defined. We return to this
issue again in Section~\ref{real}.

Here, we emphasize that, only by using our quantum hamiltonian, a description of our universe is found where it cools during its expansion, so that
the quantum state used to describe it, is very close to the lowest energy eigenstate of this hamiltonian, and has exceptionally low total entropy.
This state is far from any of the ontological states of the automaton.

\newsecl{The existence of an energy density}{convergence}

\def\intt{\mathrm{int}}
There are technical difficulties in our model that have to be considered carefully. There may be reasons to put in doubt the validity of our
derivation of a local hamilton density. When sandwiched between states whose energies differ by much more than the Planck energy, the
Baker-Campbell-Hausdorf expansion does not converge, so terminating it at any point might lead to a totally useless expression for \(\HH(\vec x)\).

The convergence of the Baker-Campbell-Hausdorff expansion can easily be examined. Consider the matrix equation \eqn{BCH} with two expansion
parameters \(s\) and \(t\),
 \be e^{sP}\,e^{tQ}\iss U(s,t)\iss e^{R(s,t)} \ , \eel{BCHst}
and its convergence upon expansion with respect to \(s\) and \(t\). The expansion of the two exponents may be expected to converge factorially. This
requires some assumption concerning the boundedness of \(P\) and \(Q\), but since \(P\) and \(Q\) represent some permutation of the states in our
deterministic model, we might assume such bounds to exist, so that the expansion of the operator \(U(s,t)\) may be assumed to converge. Now consider
the operators \(U\) and \(R\) in the complex plane of the variable \(s\), or \(t\), or some combination of both. It is in taking the logarithm of
\(U(s,t)\), so as to obtain the operator \(R(s,t)\), that cusp singularities arise. While two eigenvalues of \(R(s,t)\) may reach values that are an
integral multiple of \(2\pi i\) apart, the corresponding eigenvalues of \(U\) merge. Infinitesimal variations of \(U\) will then have singular
effects on the eigenstates of \(R\). This is where cusp singularities typical for a logarithmic function are to be expected --- and not before such a
situation is reached. As long as \(s\) and \(t\) are given such values that the eigenvalues of \(R(s,t)\) are all separated by a distance whose
absolute value is less than \(2\pi\), one expects the expansion in \(s\) and \(t\) to converge.

At first sight, this seems to be an uncomfortably small circle of convergence. However, if the fundamental unit of time is chosen to be something
comparable to the Planck time, \(10^{-44}\) sec., the radius of convergence is an energy range comparable to the Planck energy, \(10^{19}\) GeV,
which is much more than the energies involved in any conventional quantum experiment. \emph{All we have to do is limit ourselves to energies, or
energy differences, much smaller than the one dual to the fundamental time scale of the automaton.} The problem with this is, that for arbitrarily
chosen models, nothing of interest happens at time scales much longer than the typical lattice scale. This is where the hierarchy problem of
subatomic physics enters into our considerations: it is essential to have models with vastly different scales at which interesting physical phenomena
take place. Nature seems to be described by such a model, but it is notoriously difficult to produce such models from scratch, a problem known as the
\emph{hierarchy problem}.

Even if we limit ourselves to experiments where energy differences of the order of the Planck energy are not considered, there is the danger that
such energy differences are unavoidable. The automaton has an internal clock. This clock appears to violate time translation invariance for time
steps that are not an integral multiple of the clock time, or, energy might be conserved \emph{only} modulo \(2\pi\hbar/\D t\), where \(\D t\) is the
clock period. This violation might be revealed in the following way:

In conventional quantum models, one often splits the hamiltonian \(H\) into an unperturbed one, \(H_0\) and a perturbation term, \(H^\intt\). In a
deterministic theory, one might consider doing the same thing: \(H_0\) is the hamiltonian of an integrable version of the automaton, such that the
eigenstates and eigenvalues of \(H_0\) can be found exactly. Then, one introduces `exceptional' interactions that are not integrable, giving rise to
a correction \(H^\intt\). If, however, our automaton receives updates at the beat of a clock, then \(H^\intt\) may only conserve the energy modulo
\(2\pi\hbar/\D t\), and so there is the danger of energy non-conservation, which would jeopardize any application of thermodynamics, and eventually
also invalidate the perturbation procedure itself, because in the absence of a ground state perturbative effects would diverge very badly -- a
problem that could be suspected to be related to that of the BCH divergence.

Note however that such violations of energy conservation would be impossible to reconcile with general relativity, which should be one indication
that this cannot really happen. We therefore propose a different way to deal with approximate solutions. Consider the full hamiltonian \(H\)
calculated in the expression~\eqn{Hsum}. Splitting it into an ``unperturbed" part \(H_0\) and a perturbation term \(H^\intt\) should \emph{not} be
done as sketched above, but by starting with the full-fledged expression \eqn{Hsum}. This hamiltonian is exactly conserved, so that it never can
produce energy jumps.

Apparently, we are forced, at an early stage, to limit ourselves to the quantum formalism. We should \emph{not} consider perturbations that replace
an unperturbed ontological evolution law by a perturbed one. But this was precisely the perturbation that in Section~\ref{Bell} caused us
difficulties because it would have yielded classical Bell inequalities. The quantum perturbation theory leaves us with a quantum description
throughout. The ``unperturbed" system is no longer a deterministic cellular automaton but a genuine quantum world. The perturbation may seem to make
it look even more complicated, but actually only the complete hamiltonian represents our deterministic model.

The procedure that leads to a hamilton density will be far from unique. At every stage, there will be the freedom to add multiples of \(2\pi\hbar/\D
t\) to the eigenvalues of \(H\). Yet all these hamiltonians will give theories in which the cellular automaton's data evolve in exactly the same
manner as long as the time intervals are restricted to integral multiples of \(\D t\). All these theories will give different vacuum states, but only
the (physically irrelevant) phase factors will be different.

\newsecl{Can the real world be a cellular automaton? Speculations about symmetries.}{real}

The models studied so-far, cellular automata on a lattice, appear to lack several of the symmetries that we do observe in the real world. In the real
world we have Poincar\'e invariance. The cellular automaton's lattice is not even rotationally invariant. A finite value for the lattice link size
would also indicate a breakdown of translation invariance in space and in time, from a continuous symmetry into a discrete subgroup, but this may be
cured in the large distance limit. The point is that, in our models, conservation of energy is absolute, rather than a conservation modulo
\(2\pi\hbar/\D t\), so that time translation is a continuous symmetry. Similarly, momentum will be an integral of momentum density, so that also
momentum may extend to infinity instead of the first Brillouin zone of the lattice. Conceivably, for some cellular automaton models, this may lead to
continuous translation invariance instead of a discrete one; however, continuous displacement invariance, and the associated additive conservation of
momentum at all scales does not come automatically and must be proven, since all we have on a lattice is momentum conservation up to the Brillouin
zone.

It is here that we suspect models with information loss may be needed. These models lack time reversibility at the Planck scale.
As explained in earlier work,\cite{GtHbeable} this forces us to identify quantum states not with the ontological states of the
model but with equivalence classes of states; two states are equivalent is, after any amount of time, they evolve into the same
state. This leads to a quantum Hilbert space where time reversibility is restored.

Information loss may have as an effect that small-distance structures are wiped out, so that, consequently, Hilbert space will be dominated by low
momentum states only. Thus, Fock space may be predominantly occupied by sets of particles each having momenta much smaller than the Planck momentum.
Upon interactions, these momenta then have to be exactly conserved, so that we recuperate absolute momentum conservation. At this point, this is no
more than a conjecture.

\begin{figure}[h]
\begin{quotation}
 \epsfxsize=120 mm\epsfbox{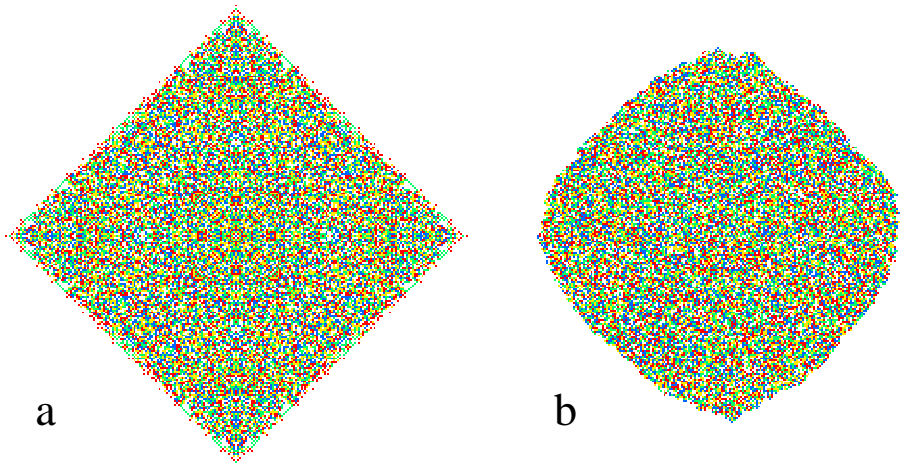}
  \caption{Sketch of a typical deterministic, time reversible cellular automaton (\(\mathbb{N}=5\))
after 120 steps. $a$) starting from initial data of a single cell being one, surrounded by empty
cells. $b$) The same automaton, starting from random initial data, and comparing with what happens
if one cell at the center is changed by one unit. One then observes an approximate rotational
invariance.}
  \label{ca.fig}\end{quotation}
\end{figure}

Once we see how this works, one can imagine that a lattice model also recuperates rotational invariance spontaneously at large
distance scales, were the lattice effects become invisible. This is not at all guaranteed; the speed by which information spreads
may well stay angle dependent, see Fig.~\ref{ca.fig}. In Fig.~\ref{ca.fig}$a$, we see how an automaton might evolve from an
initial configuration where only one of the cells is given a value different from all others. In general, this will not be
rotationally invariant\fn{Further experiments with cellular automata show that \emph{approximate} rotational invariance can arise
in many cases.}, but that is also not needed. Our vacuum state will be a superposition of many states. We can ask how a
changeable operator evolves. In Fig.~\ref{ca.fig}$b$, it is illustrated how surrounding cells respond to a modification of one
cell at \(t=0\), after 100 time steps. This often generates a pattern for the speed of information transfer that is approximately
rotationally invariant, but not quite.

Absolute invariance under rotations may again require models with information loss. Besides that, there is a curious observation to be made regarding
fermions. Anticommuting fields are a natural choice for describing discrete objects, particularly data that can only take the values 0 and 1, on a
lattice. With such fields, one can decompose these degrees of freedom into Fourier modes, which leads to fermionic particles. Now let us consider a
rotation of such Fourier modes. The lattice only allows rotations that are elements of the discrete crystal group of the lattice. Thus, it is easy to
put these fermions in a representation of these rotations. However, the number of lattice points that is transformed into one another in any
representation, is always even.\fn{This argument assumes that we only consider rotations in at least three dimensions; there, all platonic bodies
have an even number of points.} This means that these fermionic states must be in even dimensional representations of the rotation group. This may
well explain why the spin of these one-particle states must be half-odd integer.

The lattice crystal group has only low-dimensional representations, so that single particles can only have low values for both \(L\) (orbital angular
momentum) and \(S\) (spin). Again, adding the feature of information loss may help us to add larger values of orbital angular momentum by only
allowing states with relatively low values of momentum.

As a next step, one can ask how such a model could also develop invariance under Lorentz transformations, by employing similar mechanisms; a feature
that could be helpful here is the fact that the speed of information transfer, to be interpreted as the speed of light, is clearly limited in all
directions. In addition to the generators for rotations, we now also need to identify generators for Lorentz transformations. Again, these could be a
mix of beables and changeables.

As yet, the question whether such procedures can be implemented to obtain a cellular automaton with Lorentz invariance is even more difficult to
address than the other symmetries, translation and rotation. For this reason, we do not know whether our cellular automaton can generate particles
with a dispersion law (the relation between wave number and frequency, or equivalently, the relation between momentum and energy) that is anything
resembling the usual Lorentz invariant one. Presumably, very special conditions have to be specified for the continuum limit of an automaton to
display a symmetry as delicate as Lorentz invariance. Since we suspect that such Lorentz transformations will connect beables with changeables,
Lorentz invariance will not be easy to perceive at the classical level.

Even more perplexing is the fact that our world appears to be invariant under general coordinate transformations. Either this should be ascribed to a
glassy structure of our space-time lattice, or else this symmetry is also due to our primitive quantization procedure: a symmetry connecting beables
with changeables.

The lattice length scale is to be expected to be close to the Planck scale. There are two reasons for this assumption. One is that the emergence of
Lorentz symmetry and general invariance are both difficult to understand, so that most likely they come together; a cubic lattice could well serve at
distance scales different from the Planck length but would be difficult to reconcile with Lorentz invariance separately. A more compelling argument
for the lattice scale to coincide with the Planck scale is the existence of black holes and the phenomenon of Hawking radiation, which could be
derived by standard quantum field theoretical techniques. Thermodynamics then tells us right away that the entropy of a black hole is equal to
\(\quart\ \times\) the horizon area in Planck units, and this is a direct measure for the total number of states a black hole can be in, so that the
discrete cellular degrees of freedom indeed have to be counted at Planckian dimensions.

Although this gives us a huge number of states for small systems, a large black hole seems to show far too few states, since their number is counted
on the horizon and not in the bulk. The fact that there are so few states is to be interpreted as being due to the fact that we count equivalence
classes rather than ontological states; indeed, this is our most compelling argument for information loss.

Clearly, we are not even close to describing realistic cellular automata equations for the real world, but this was not our ambition in this paper.
Here, we restrict ourselves to the observation that time-reversible cellular automata have some properties that could explain the quantum mechanical
nature of our world, even if we do not understand all the details, and we have not yet been able to demonstrate all physically desirable symmetry
patterns.

\newsecl{Conclusion}{conclusion} 

The experimental setup for which Bell's inequalities can be tested relies not only on our ability to distinguish quantum entangled states
from classically entangled states, but also on an assumption concerning the way this quantum entangled state is being produced. An atom of
a species that is well studied, is assumed to produce two photons that are entangled in the same way as they always are, for instance by
having total spin zero. Think of an \(S\) state decaying into a lower \(S\) state by the emission of two photons.

If Alice measures one component of the spin, Bob can measure another component and any knowledge of what Alice has found will lead to
apparent conflicts. The conflicts only arise from our assumed knowledge of the fact that the total spin is zero. If the effect of a change
in Alice's settings would be that the total spin of the photons might change into a superposition of spin zero and spin two then there
would be a new situation. In a deterministic underlying theory this could happen. The point is that the phases of all states are, in a
sense, introduced by hand; the phases in the amplitudes in terms of the ontological basis of Hilbert space have no physical meaning at all.
A two photon state with total spin zero is not an eigenstate of the beables of the theory. Therefore, its description relies on our
conventions concerning the phases of this state. If we may assume that the initial state of the universe is a quantum entangled state, then
any operation of a changeable operator in quasar \(A\) may require a careful rearrangement of these artificial phase factors, even for
objects in our system \(S\), that is far and spacelike separated from the quasar. This is why a direct measurement of the total spin of the
photons might always give zero, but the operation of the changeable at quasar \(A\) might nevertheless affect the phases in such a way that
the total spin would seem to be two, unless we allow for a state that also would modify the outcome of Bob's measurement.

Note that the state with total spin two would not be forbidden by any observations of the total energy and entropy at the time of
measurement, or at \(t=t_1\). The state is legal, and it can be extrapolated back to the past using Schr\"odinger's equation.
However, since the spin two configuration cannot have a past in terms of a single particle \(\e\) as the one prepared by the
experimentalists, this past reconstruction of the state clashes with conventional thermodynamics; the entropy will increase
towards the past, while it should have decreased according to conventional wisdom. This, we suspect, is what our entangled state
would look like if we attach incorrect phase factors to the amplitudes in the ontological basis.

The claim that a quantum theory can be distinguished from a classical theory using Bell's inequalities makes use of the laws of
thermodynamics in a special way: it is assumed that our environment is very close to the state we call vacuum. In this vacuum state, many
phase angles are assumed to take very special values (otherwise, we have some highly excited, and therefore `very improbable' state). It is
these phase angles that are man-made; modifying them has no physically observable effects other than replacing all states that we use to
describe the situation, by states we like to regard as improbable (since entropy is far higher than in the most natural description). In
our hypothesis, non-local modifications of these phase factors are necessary for a consistent quantum mechanical description, in terms of
low energy/entropy states, but they do not have any direct physical meaning.

Remarkably, an essential ingredient of our hypothesis concerns the fact that the universe evolves, so as to generate initial states with
not only low energy content but also low entropy. Energy and entropy are defined only for the quantum states, not for the ontological
states. For the quantum states, we need the phase factors even if they are man-made artifacts.

Thus, we found sufficient motivation to proceed along the path chosen here: take some classical cellular automaton, use quantum operators
to describe its time evolution, write the hamiltonian as an integral of an Hamilton density and treat the resulting theory as a
full-fledged quantum field theory. The quantum states that serve as its basis will have to be interpreted entirely in the spirit of the
Copenhagen doctrine. As such, there is no reason to expect these states to obey Bell's inequalities.

\begin{figure}[h]
\begin{quotation}
 \epsfxsize=105 mm\epsfbox{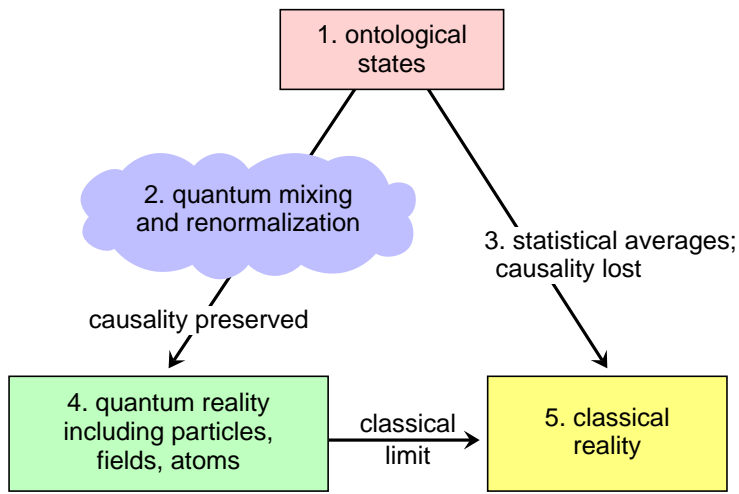}
  \caption{The relation between ``ontological reality" at the Planck scale (1), quantum reality at the atomic scale (4) and classical
reality (5). The arrow along step 2, using quantum superposition, is mathematically exact, preserving strict causality, but no
longer refers to ontological states. Step 3, involving large numbers, preserves ontological reality, but looses strict causality
in describing nature's laws.}
  \label{onto.fig}\end{quotation}
\end{figure}

The picture we arrive at, is sketched in figure \ref{onto.fig}. When using the renormalization group in passing from the Planck
scale to the atomic scale, one is forced to apply quantum superposition, arriving at a description of effective objects as
quantum states, including particles as well as fields, described by non-commuting operators. The expectation values of averages
in the classical limit probably coincide with the ones obtained directly from averaging the ontological states at the Planck
scale, so that the classical limit again describes classical reality, but he price paid is that strictly causal \emph{effective}
laws of nature cannot be formulated for these classical objects.

It is important to emphasize that, at stage 4, none of the particles and fields are ontological realities. They are the elements
of a Hilbert space needed if one insists in preserving exact effective laws of nature at that scale. The classical objects seen
at step 5 are statistical averages of large numbers of objects at the Planck scale (1), but strict causality was lost in the
statistical averaging procedure. Having no causal `effective' laws makes it impossible to describe the classical world at stage 5
entirely in terms of classical `logic': the events one observes appear to be `inexplicable'. Only `quantum logic', as we have
learned at stage 4, appears to work.

So-far, our conclusions do take the form of an hypothesis. Further verification may come if we find more realistic cellular automaton
models, generating more interesting quantum field theories with a closer resemblance to the Standard Model, for instance by being exactly
or approximately Lorentz invariant.

The need for information loss at the Planck scale was not touched upon much in this paper, but the author does believe information loss to
be important. The distinction between low entropy states and high entropy ones as mentioned may well require this ingredient, but it is
still not quite understood. This is one of the reasons why we insist of our conclusions to be phrased with care. Further investigations are
much needed.

\end{document}